\documentclass[10pt,conference]{IEEEtran} 
\usepackage{cite}
\usepackage{amsmath,amssymb,amsfonts}
\usepackage{algorithm}
\usepackage{algpseudocode}

\usepackage{graphicx}
\usepackage{textcomp}
\usepackage{xcolor}
\def\BibTeX{{\rm B\kern-.05em{\sc i\kern-.025em b}\kern-.08em
    T\kern-.1667em\lower.7ex\hbox{E}\kern-.125emX}}
\begin{document}

\title{LLM-Guided Ansätze Design for Quantum Circuit Born Machines in Financial Generative Modeling \\
{}

}

\author{\IEEEauthorblockN{1\textsuperscript{st} Yaswitha Gujju}
\IEEEauthorblockA{\textit{Computer Science} \\
\textit{ University of Tokyo}\\
Tokyo, Japan \\
yaswitha-gujju@g.ecc.u-tokyo.ac.jp}
\and
\IEEEauthorblockN{2\textsuperscript{nd} Romain Harang}
\IEEEauthorblockA{\textit{Computer Science} \\
\textit{ University of Tokyo}\\
Tokyo, Japan \\
romain-harang@g.ecc.u-tokyo.ac.jp}
\and
\IEEEauthorblockN{3\textsuperscript{nd} Tetsuo Shibuya}
\IEEEauthorblockA{\textit{Computer Science} \\
\textit{ University of Tokyo}\\
Tokyo, Japan \\
tshibuya@hgc.jp}
}


\maketitle
\IEEEpeerreviewmaketitle

\begin{abstract}
Quantum generative modeling using quantum circuit Born machines (QCBMs) shows promising potential for practical quantum advantage. However, discovering ansätze that are both expressive and hardware-efficient remains a key challenge, particularly on noisy intermediate-scale quantum (NISQ) devices. In this work, we introduce a prompt-based framework that leverages large language models (LLMs) to generate hardware-aware QCBM architectures. Prompts are conditioned on qubit connectivity, gate error rates, and hardware topology, while iterative feedback—including Kullback–Leibler (KL) divergence, circuit depth, and validity—is used to refine the circuits. We evaluate our method on a financial modeling task involving daily changes in Japanese government bond (JGB) interest rates. Our results show that the LLM-generated ansätze are significantly shallower and achieve superior generative performance compared to the standard baseline when executed on real IBM quantum hardware using 12 qubits. These findings demonstrate the practical utility of LLM-driven quantum architecture search and highlight a promising path toward robust, deployable generative models for near-term quantum devices.

\end{abstract}

\begin{IEEEkeywords}
Quantum generative modeling, Quantum circuit Born machine (QCBM), Financial time series, Quantum hardware, Noise mitigation, Large language models, optimization
\end{IEEEkeywords}

\section{Introduction}

Quantum machine learning (QML) \cite{b18,b19,b20} promises to unlock the capabilities of near-term quantum devices by offering novel approaches to data-driven tasks. Among various QML paradigms, generative modeling using parameterized quantum circuits—such as the quantum circuit Born machine (QCBM)—has emerged as a compelling direction. QCBMs leverage the Born rule and repeated measurements to learn target data distributions by preparing quantum states through a variational ansatz.

Previous works~\cite{b8} have demonstrated that QCBMs can effectively model both synthetic and real-world datasets, indicating potential quantum advantage in generative tasks. However, practical deployment remains challenging due to hardware noise and the limited adaptability of standard ansatz templates. Transpilation often increases circuit depth substantially, exacerbating noise sensitivity and hindering execution on current quantum processors.

In this work, we propose a hardware-aware quantum architecture search framework leveraging large language models (LLMs). By prompting the LLM with hardware-specific constraints—such as qubit connectivity and error rates—and providing iterative feedback based on generative performance metrics, we enable the synthesis of compact, efficient ansatzes tailored for real devices. Applying this approach to financial data modeling, we show that LLM-generated circuits outperform standard baseline in both noise robustness and fidelity on real quantum hardware. Our results demonstrate the promise of LLMs in automating adaptive quantum circuit design, advancing the practicality of quantum generative modeling on near-term devices.

\section{Background}
\subsection{Quantum Circuit Born Machine (QCBM)}

QCBMs are a class of quantum generative models capable of learning arbitrary probability distributions from data. Given a data set $Y \in \{0,1\}^n$ with an underlying distribution $p_Y$, we define a parameterized quantum circuit $U(\theta)$, comprised of gates, some of which are parametrized by $\theta$, that prepares the quantum state.
\[
|\psi_\theta\rangle = U(\theta) |0\rangle^{\otimes n}.
\]
Measurements on this state yield bitstrings $X \in \{0,1\}^n$, which define a learned distribution $p_X$. The training objective is to optimize the parameters $\theta$ such that $p_X$ closely approximates $p_Y$. To model arbitrary finite distributions, the data is encoded into binary strings. For real-valued or continuous data, discretization is performed via uniform binning based on the bit-depth determined by $n$.

Choosing an appropriate loss function is crucial for effective training. While Kullback-Leibler (KL) divergence~\cite{b17} is a common metric for comparing distributions, it often leads to poor convergence in QCBMs due to instability in gradient estimation ~\cite{b8}. As an alternative, the Maximum Mean Discrepancy (MMD) \cite{b22} is widely adopted as an implicit loss function:

\[
\begin{aligned}
\mathcal{L}_{\text{MMD}}(p_X, p_Y) =\ 
& \mathbb{E}_{x,x' \sim p_X}[\kappa(x,x')] \\
& + \mathbb{E}_{y,y' \sim p_Y}[\kappa(y,y')] \\
& - 2 \mathbb{E}_{x \sim p_X, y \sim p_Y}[\kappa(x,y)],
\end{aligned}
\]

where $\kappa$ is a kernel function (typically the radial basis function, gaussian mixture kernel, or similar), treated as a hyperparameter.

\begin{figure*}
    \centering
    \includegraphics[width=0.85\linewidth]{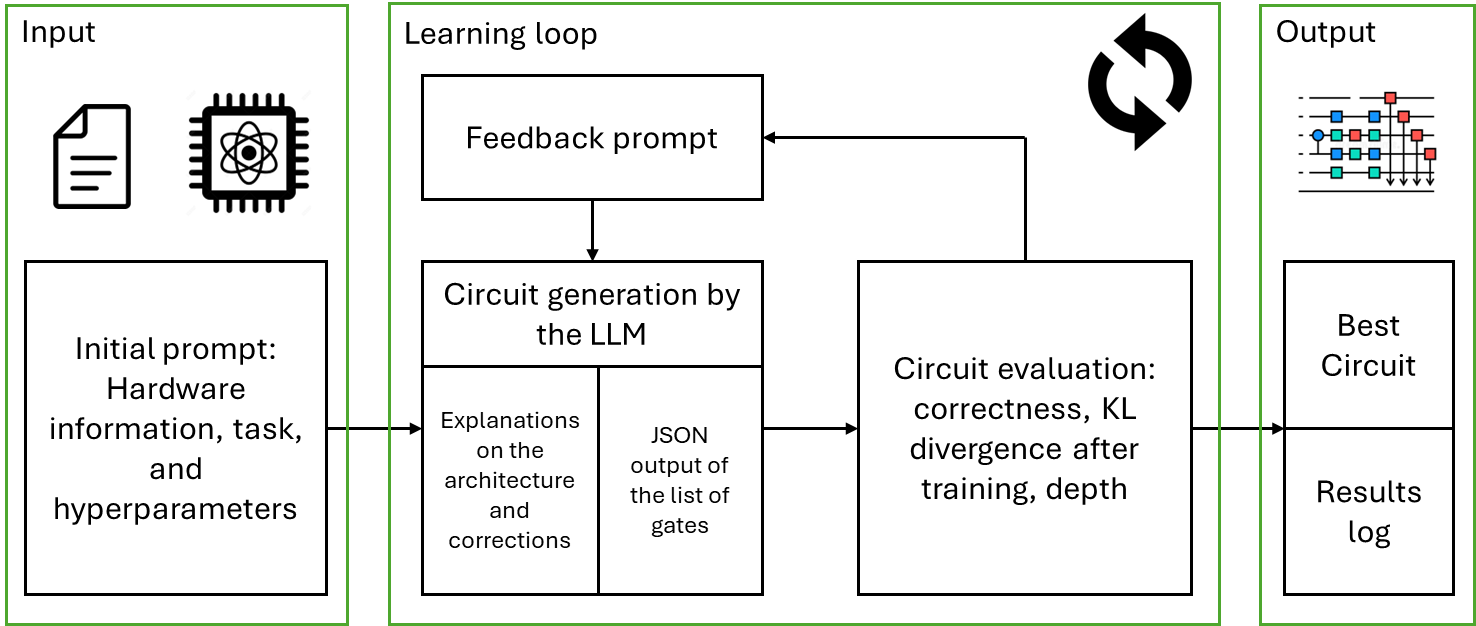}
    \caption{Overview of our experimental setup}
    \label{fig:pipeline}
\end{figure*}

\subsection{Previous Works}
The use of LLMs for quantum circuit design is a recent and rapidly developing research direction. Typically, the LLM is treated as a black box that takes a set of instructions (a prompt) and generates textual outputs conditioned on that prompt. For example, the work in~\cite{b10} introduced a supervised learning approach to ground-state problems, generating task-specific circuits that were validated through simulation. Similarly,~\cite{b9} proposed an LLM-driven ansatz selection strategy that incorporates experimental feedback, though their method remains largely conceptual and lacks detailed implementation. These approaches often rely on fixed circuit templates and static internal models, which can limit adaptability across tasks and hardware. In contrast,~\cite{b11} developed a modular, prompt-based system for quantum feature map generation, combining LLM-driven synthesis, empirical validation, and iterative refinement through literature search and performance benchmarks. This work marks progress toward fully autonomous circuit discovery.

Separately, prior studies on QCBMs~\cite{b1,b8} have demonstrated their capability to model simple datasets such as bars-and-stripes and interest rate time series. However, these studies are typically limited to simulators, with minimal comparisons to classical baselines or deployment on real hardware.

\section{Our Method}

Our approach leverages LLM to autonomously generate and iteratively refine QCBM ansatz tailored to both the target generative modeling task and the constraints of the available quantum hardware. Starting from an initial prompt that encodes hardware profile (including qubit count, connectivity, and error rates), the LLM proposes a  candidate quantum circuit. This candidate is evaluated according to multiple metrics such as the KL divergence, circuit validity, and depth of the circuit to guide an iterative feedback loop.
The KL divergence between two distributions \( P \) and \( Q \) is defined as
\[
D_{\mathrm{KL}}(P \,\|\, Q) = \sum_{x} P(x) \log \left( \frac{P(x)}{Q(x)} \right),
\]
where the sum is taken over all \( x \) such that \( P(x) > 0 \).

In this work, we compute the reverse KL divergence \(D_{\mathrm{KL}}(Q \parallel P)\), where \(Q\) is the model distribution and \(P\) is the true data distribution, as it better reflects the model’s tendency to avoid assigning probability to unlikely outcomes. This choice also improves numerical stability on noisy quantum hardware by focusing on the support of \(Q\), making it a more practical metric for evaluating generative performance. Henceforth, we refer to reverse KL divergence as simply KL divergence.

By providing performance feedback to the LLM in subsequent prompts, the system nudges the circuit design toward increasingly optimal ansatz configurations, balancing expressivity and hardware feasibility. Figure \ref{fig:pipeline} provides a summary of our proposed approach.

\subsection{Prompts}
\paragraph{Initialization} To ensure that the generated circuits comply with the hardware constraints—such as qubit connectivity and native basis gates—we use the LLM to generate a new circuit at each round, taking the specific hardware configuration into account. Additionally, we preferentially select qubits with lower error rates to avoid the need for further transpilation. This is important because transpilation typically increases circuit depth, which in turn degrades performance on real quantum hardware due to the accumulation of noise.

\paragraph{Feedback and Architecture Evolution}
In each round, we provide the LLM with performance metrics of the previously generated circuit, including the KL divergence and the circuit depth after training. We additionally specify in the prompt that the LLM should prioritize reducing the KL divergence over increasing circuit depth, in order to resolve the trade-off between these two competing objectives. A maximum allowable depth is also provided to prevent the LLM from generating circuits that may become infeasible to run on the real quantum hardware.
 The training itself is performed using the MMD loss. Additionally, we include a boolean variable \texttt{validity}, which indicates whether the generated circuit adheres to the hardware constraints such as allowed basis gates and qubit connectivity. This serves to ensure that the circuits remain executable on real quantum hardware. Based on these inputs, the LLM decides whether to prune the current circuit or append new layers, with the goal of reducing the KL divergence and improving model performance.



\begin{figure*}[!ht]
    \centering
    \includegraphics[width=0.49\textwidth]{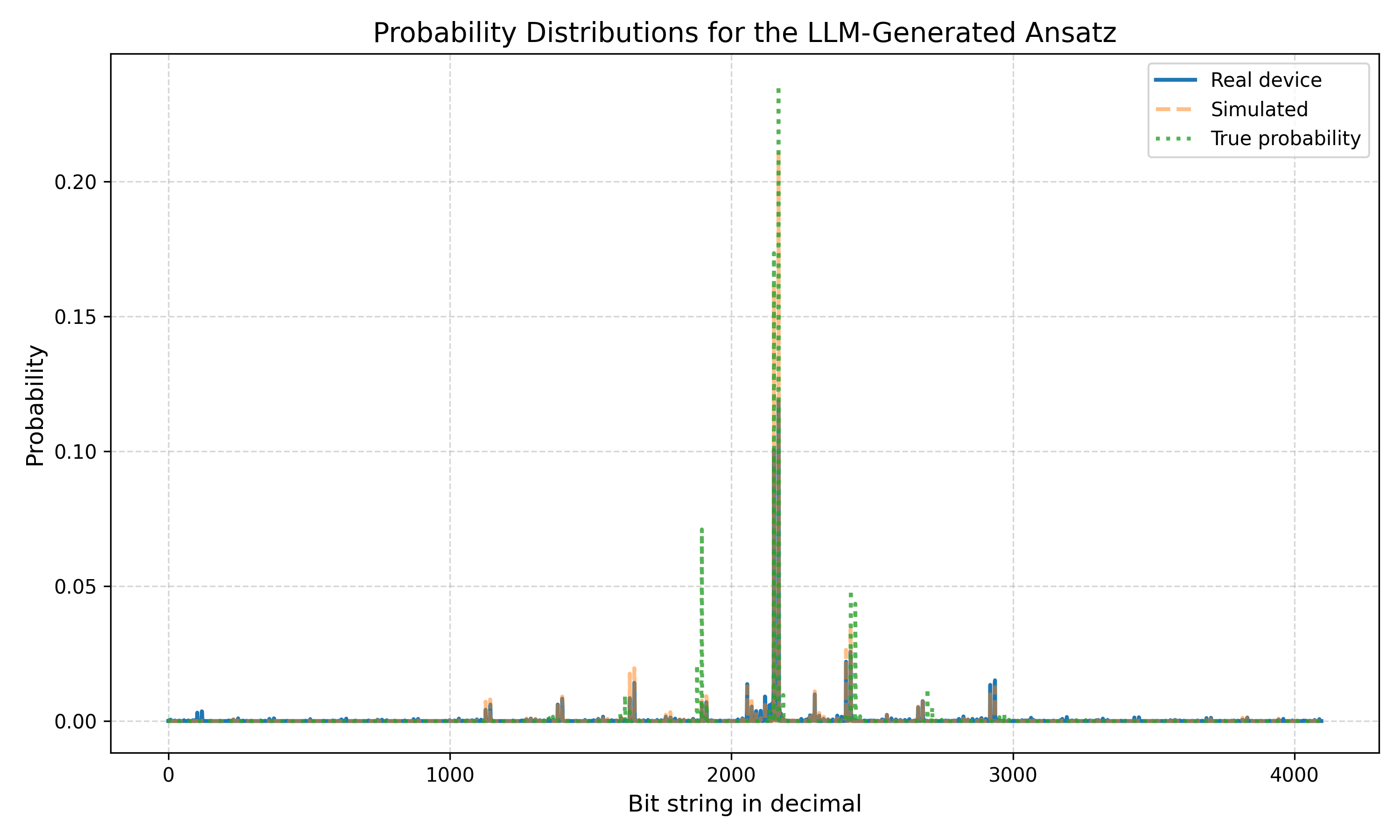}
    \includegraphics[width=0.49\textwidth]{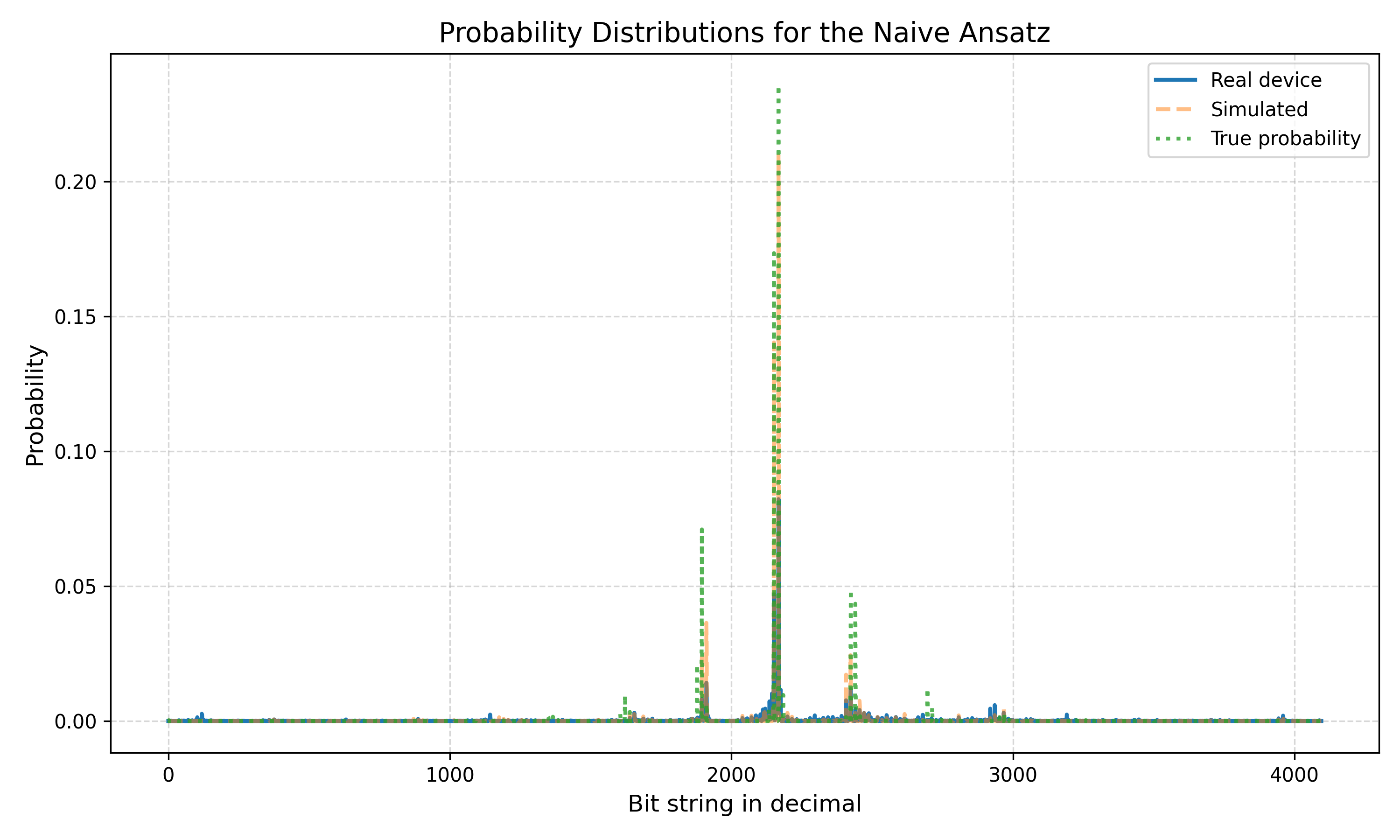}
 \caption{Probability distributions of the true data (green) and those generated by the QCBM using both the LLM-generated and standard (naive) ansatz, evaluated on a noiseless simulator (orange) and real quantum hardware (blue). The left panel shows the distribution of the LLM-generated ansatz while the right panel shows the naive ansatz distribution.}

    \label{fig:naivehist}
\end{figure*}
\section{Experiments}

\subsection{Dataset and Preprocessing}
The generative modeling task in this work focuses on interest rate time series. Following~\cite{b1}, we use Japanese Government Bond (JGB) rates instead of more commonly used foreign exchange datasets, as interest rates exhibit stronger temporal and cross-maturity correlations due to the yield curve structure. The dataset is adopted from the publicly available Ministry of Finance Japan, comprising daily interest rates for three JGB maturities (5, 10, and 20 years) spanning from January 4, 2000, to February 28, 2025. In total, 12 qubits are used, allocating 4 qubits per JGB maturity.

To prepare the input for quantum circuit processing, we follow the binary encoding scheme described in~\cite{b1}, which applies uniform quantization at a fixed bit resolution. Each real-valued input $x$ is mapped to its $n$-bit binary encoding as:
\[
(x)_2 = \left\lfloor \frac{(2^n - 1)(x - x_{\min})}{x_{\max} - x_{\min}} \right\rfloor,
\]
where $x_{\min}$ and $x_{\max}$ are the dataset's minimum and maximum values, respectively. The inverse mapping is given by:
\[
(x)_{10} = x_{\min} + \frac{(x)_2 (x_{\max} - x_{\min})}{2^n - 1}.
\]

\subsection{Training Setup and Hardware}

Training is performed on the noiseless \texttt{aer\_simulator} backend in Qiskit~\cite{b12}, executing 10{,}000 measurement shots per circuit evaluation. The Adam optimizer~\cite{b23} with a learning rate of 0.1 is used, along with stochastic gradient descent over 30 epochs. Mini-batches of size 1{,}000 are randomly sampled for gradient estimation, while the loss is computed over the full training set of 6{,}164 samples.

We adopt the MMD loss with a Gaussian kernel of bandwidth $\sigma = 3$, consistent with prior work~\cite{b8}. After training, the circuit is executed directly on the 12-qubit \texttt{ibm\_fez} quantum processor without requiring transpilation, due to hardware-aware initialization via prompt engineering. Gradients are estimated using the parameter-shift rule, where partial derivatives with respect to each parameter are computed individually~\cite{b24}. Consequently, the computational cost scales linearly with the number of parameters, making simpler circuits faster to train. This efficiency enables convergence analysis of LLM-generated circuits over extended training durations, typically spanning 50 to 100 epochs.

\subsection{Baseline Model}

As a baseline, we employ the \texttt{TwoLocal} ansatz from Qiskit as the QCBM architecture. It consists of alternating parameterized rotation layers (\texttt{RX}, \texttt{RZ}) and entanglement layers composed of \texttt{CZ} gates arranged in a linear (ring) topology. The circuit is implemented using the same basis gates as the target hardware and is configured with 18 repetitions, resulting in a depth of approximately 85 to ensure competitive expressivity. 

The baseline model uses the same number of qubits and training configuration as our method, allowing for a fair comparison of generative performance.

\subsection{LLM Configuration}
Quantum feature map design leverages OpenAI's LLMs, primarily GPT-4.1 and GPT-4o (latest as of May 2025) \cite{b21}. Due to formatting and reasoning consistency, GPT-4.1 was selected for generating all experimental results. The temperature parameter was kept at default, and prompt engineering included explicit reasoning steps to ensure adherence to output formats. Models from the o3 series were excluded due to lower suitability under our constraints.

\begin{figure}[!ht]
    \centering
    \includegraphics[width=1\linewidth]{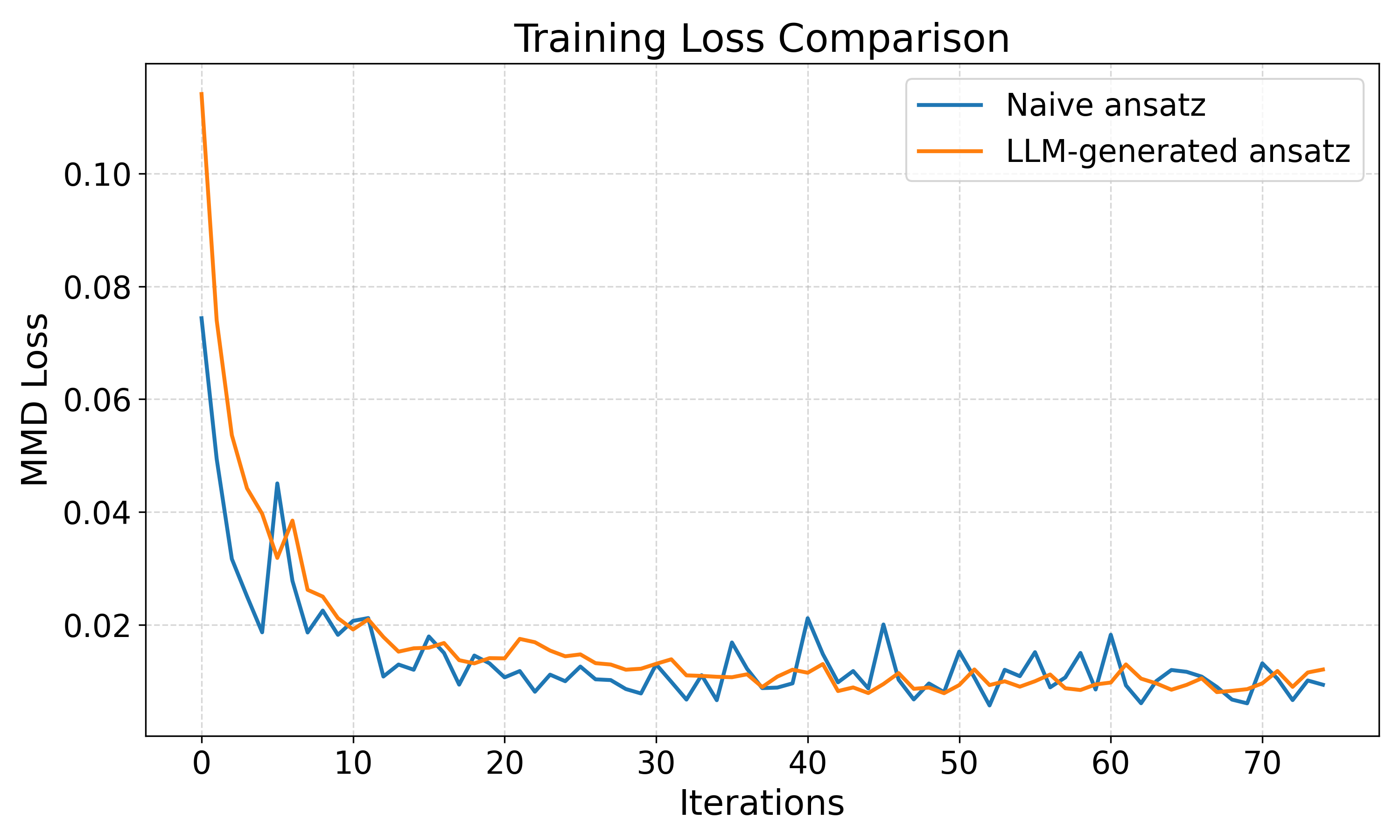}
    \caption{MMD loss during training for the ansatz generated using the LLM compared to the naive ansatz over 15 epochs, the values converge around 0.006}
    \label{fig:losshistory}
\end{figure}

\begin{table}[ht]
\centering
\caption{KL Divergence Comparison of LLM-Generated and TwoLocal Ansatzes}
\label{tab:results}
\begin{tabular}{|l|c|c|}
\hline
\textbf{Ansatz} & \textbf{Depth} & \textbf{KL Divergence} \\
\hline
\multicolumn{3}{|c|}{\textit{Simulator (Noiseless, Qiskit Aer)}} \\
\hline
LLM-generated         & 28  & 3.67 \\
TwoLocal              & 85  & 3.10 \\
\hline
\multicolumn{3}{|c|}{\textit{Real Hardware (ibm\_fez, 12 Qubits)}} \\
\hline
LLM (no EM)           & 28  & $7.37 \pm 0.11$ \\
LLM (with EM)         & 28  & $6.91 \pm 0.13$ \\
TwoLocal (no EM)      & 85  & $9.32 \pm 0.06$ \\
TwoLocal (with EM)    & 85  & $8.92 \pm 0.07$ \\
\hline
\end{tabular}
\end{table}

\section{Results}

\subsection{Simulator Results}

We evaluate QCBM performance on the noiseless \texttt{aer\_simulator} backend in Qiskit, comparing our LLM-generated ansatz to the standard \texttt{TwoLocal} ansatz. Both achieve similar performance in terms MMD and KL divergence. However, their circuit structures differ significantly: the \texttt{TwoLocal} ansatz has a depth of 85 and more parameters, while the LLM-generated circuit is more compact with a depth of 28. The KL divergence is $3.67$ for the LLM-generated ansatz and $3.1$ for the \texttt{TwoLocal} baseline. Despite comparable generative performance, the lower depth of the LLM-generated circuit makes it more suitable for near-term hardware.

\subsection{Real-Device Results}

To assess hardware performance, we execute the trained circuits on the 12-qubit \texttt{ibm\_fez} quantum processor using Qiskit's \texttt{Sampler} primitive. We apply measurement error mitigation (EM) via the \texttt{Mthree} package~\cite{b13} to correct readout errors and improve fidelity.

The LLM-generated circuit's KL divergence increases to $7.37 \pm 0.11$ without EM and $6.91 \pm 0.13$ with EM. The \texttt{TwoLocal} circuit performs worse under noise, reaching $9.32 \pm 0.06$ without EM and $8.92 \pm 0.07$ with EM. These results demonstrate the superior noise resilience of the LLM-generated circuit, especially when combined with basic error mitigation.

\paragraph{Post-selection and Noise Effects}

To further analyze the impact of noise, we apply post-selection on the best real device output based on valid bitstrings and examine the resulting output distributions. Notably, we observe a substantial improvement in KL divergence after post-selection: for both the LLM-generated ansatz and the standard \texttt{TwoLocal} ansatz, the KL divergence decreases to approximately 0.78 and 0.68, respectively. This stark contrast highlights the effect of noise on generative modeling quality and highlights the importance of post-processing techniques such as error mitigation and post-selection when deploying quantum generative models on real hardware.

\section{Discussion and Limitations}
 While LLMs effectively guide ansatz generation and consistently reduce the MMD loss per iteration, they sometimes produce invalid or unsupported circuit components, especially after 10--20 iterations. This highlights challenges in model robustness and error handling that warrant future work.

Additionally, training deep QCBM is computationally demanding and further complicated by noise in current quantum hardware, necessitating effective error mitigation. Incorporating hardware details such as qubit connectivity and error rates into LLM prompts reduces transpilation overhead and controls circuit depth, improving the feasibility of deploying generated ansatz on real devices---addressing a common limitation in prior approaches and shows significant performance improvements in terms of hardware robustness in comparison to naive ansatz. 

Overall, our method produces lightweight, expressive circuits better suited for noisy, near-term quantum hardware. Exploring diverse circuit templates as prompt examples may further enhance the LLM’s ability to generate unique and valid circuits.


\begin{thebibliography}{00}
\bibitem{b1} Makarski, Mathis, et al. "Circuit Design based on Feature Similarity for Quantum Generative Modeling." arXiv preprint arXiv:2503.11983 (2025)
\bibitem{b8} Rudolph, Manuel S., et al. "Trainability barriers and opportunities in quantum generative modeling." npj Quantum Information 10.1 (2024): 116.
\bibitem{b9} Ueda, Kento, and Atsushi Matsuo. "Optimizing Ansatz Design in Quantum Generative Adversarial Networks Using Large Language Models." arXiv preprint arXiv:2503.12884 (2025).
\bibitem{b10} Nakaji, Kouhei, et al. "The generative quantum eigensolver (GQE) and its application for ground state search." arXiv preprint arXiv:2401.09253 (2024).
\bibitem{b11} Sakka, Kenya, Kosuke Mitarai, and Keisuke Fujii. "Automating quantum feature map design via large language models." arXiv preprint arXiv:2504.07396 (2025).
\bibitem{b12} Javadi-Abhari, Ali, et al. Quantum Computing with Qiskit. arXiv:2405.08810, arXiv, 15 May 2024. arXiv.org, arxiv.org/abs/2405.08810.
\bibitem{b13} Nation, Paul D., et al. "Scalable mitigation of measurement errors on quantum computers." PRX Quantum 2.4 (2021): 040326.
\bibitem{b14} Liu, Jin-Guo, and Lei Wang. "Differentiable learning of quantum circuit born machines." Physical Review A 98.6 (2018): 062324.
\bibitem{b15} Benedetti, Marcello, et al. "A generative modeling approach for benchmarking and training shallow quantum circuits." npj Quantum information 5.1 (2019): 45.
\bibitem{b16} Ackley, David H., Geoffrey E. Hinton, and Terrence J. Sejnowski. "A learning algorithm for Boltzmann machines." Cognitive science 9.1 (1985): 147-169.
\bibitem{b17} Kullback, Solomon, and Richard A. Leibler. "On information and sufficiency." The annals of mathematical statistics 22.1 (1951): 79-86.
\bibitem{b20} Cerezo, Marco, et al. "Challenges and opportunities in quantum machine learning." Nature computational science 2.9 (2022): 567-576.
\bibitem{b18} Gujju, Yaswitha, Atsushi Matsuo, and Rudy Raymond. "Quantum machine learning on near-term quantum devices: Current state of supervised and unsupervised techniques for real-world applications." Physical Review Applied 21.6 (2024): 067001.
\bibitem{b19} Biamonte, Jacob, et al. "Quantum machine learning." Nature 549.7671 (2017): 195-202.
\bibitem{b21} Islam, Raisa, and Owana Marzia Moushi. "Gpt-4o: The cutting-edge advancement in multimodal LLM." Authorea Preprints (2024).
\bibitem{b22} Gretton, Arthur, et al. "A kernel two-sample test." The Journal of Machine Learning Research 13.1 (2012): 723-773.
\bibitem{b23} Kingma, Diederik P., and Jimmy Ba. "Adam: A method for stochastic optimization." arXiv preprint arXiv:1412.6980 (2014).
\bibitem{b24} Wierichs, David, et al. "General parameter-shift rules for quantum gradients." Quantum 6 (2022): 677.
\end{thebibliography}
\end{document}